\documentclass[11pt]{article}
\begin{document}
\begin{flushright}
SJSU/TP-04-25\\
July 2004\end{flushright}
\vspace{1.7in}
\begin{center}\Large{\bf On PSI-complete and PSI{\em R}-complete 
                     measurements}\\
\vspace{1cm}
\normalsize\ J. Finkelstein\footnote[1]{
        Participating Guest, Lawrence Berkeley National Laboratory\\
        \hspace*{\parindent}\hspace*{1em}
        e-mail: JLFINKELSTEIN@lbl.gov}\\
        Department of Physics\\
        San Jos\'{e} State University\\San Jos\'{e}, CA 95192, U.S.A
\end{center}
\begin{abstract}
I construct a POVM which has $2d$ rank-one elements and which is 
informationally complete for generic pure states in $d$ dimensions,
thus confirming a conjecture made by Flammia, Silberfarb, and Caves
(quant-ph/0404137).  I show that if a rank-one POVM is required to
be informationally complete for {\em all} pure states in $d$
dimensions, it must have at least $3d-2$ elements.  I also show that,
in a POVM which is informationally complete for all pure states in 
$d$ dimensions, for any vector there must be at least $2d-1$ POVM elements
which do not annihilate that vector. 
\end{abstract}
\newpage
\section{Introduction}
Consider the following situation: you are given many copies of a quantum
system; you know they are all in the same state, but you don't know which 
state that is, and you want to perform measurements in order to find out.
If the statistics of the outcome of these measurements are sufficient to
uniquely identify the state, the measurements are called ``informationally
complete'' \cite {IC} (I-complete).  In this note I will present some 
results for a special case of this situation, in which you know that the 
system is in a pure state, but you don't know in which pure state.

A set of measurements can be considered equivalent to a single
``generalized'' measurement, which is described by a positive 
operator-valued measure (POVM) \cite {POVM}.  I will denote elements
of a POVM as $E_i$; they are positive operators which satisfy 
$\sum_{i}E_{i} = I$, and if the state of the system is denoted as
$\rho $, then the probability of the ith outcome is given by
Tr($\rho E_i$).  For a pure state $\rho = |\psi \rangle \langle \psi |$, 
that probability is the expectation value $\langle \psi |E_{i}|\psi \rangle$.
 
Pure state I-complete POVMs have been discussed
in a recent article by Flammia, Silberfarb, and Caves (Ref.\ \cite{FSC},
hereafter FSC).  I will adopt their definition, which is as follows:

{\bf Definition (PSI-completeness).}  A pure-state informationally
complete (PSI-complete) POVM on a finite-dimensional quantum system is
a POVM whose outcome probabilities are sufficient to determine any
pure state (up to a global phase), except for a set of pure states that
is dense only on a set of measure zero.

\vspace{0.5cm}
Let $d$ denote the (finite) dimension of the Hilbert space for our quantum
system.  FSC show that any PSI-complete POVM must have at least $2d$
elements; this, together with their construction of an example 
(see also Ref.\ \cite {W})
that does in fact have $2d$ elements, shows that the minimal number of
elements of a PSI-complete POVM for a system with $d$ dimensions is indeed 
$2d$.
FSC also conjecture that, for a PSI-complete POVM whose elements are all
of rank one, the minimal number would still be $2d$.  In the next section 
of this note I will confirm that conjecture by displaying a rank-one 
PSI-complete POVM with $2d$ elements.  

The definition of PSI-completeness given above  allows there to be pure 
states which cannot be identified uniquely by the expectation values of the 
POVM elements, but it does demand that any such states be confined to a 
set of measure zero.  This means that, if a pure state were selected at 
random, then with probability one it {\em would} be uniquely identified.  
Of course,
in practice we could never measure those expectation values with infinite
precision, which means that we should not expect to identify the state
with infinite precision.  One might hope that, if we knew the expectation
values to a good approximation, we would then be able, with probability
one, to identify the 
state to a good approximation, in the sense that, (outside of a set of measure
zero) any two pure states which
were compatible with the same imprecisely-known set of expectation
values would necessarily be close together (in, say, the Hilbert-space norm).
However, PSI-completeness does not guarantee this.  Consider two distinct
states $|\psi _{a}\rangle$ and  $|\psi _{b}\rangle$ which were both 
compatible with the
same precisely-given set of expectation values; then imprecisely-known
expectation values would be compatible with  states sufficiently close 
to   $|\psi _{a}\rangle$ and also to states sufficiently close to 
$|\psi _{b}\rangle$.
Since the set of states sufficiently close to $|\psi _{a}\rangle$ 
or to $|\psi _{b}\rangle$ has finite
measure, there would be a (small but) finite probability that we  would not
be able to know if the state was close to $|\psi _{a}\rangle$ or to 
$|\psi _{b}\rangle$.

We could strengthen the definition of PSI-completeness by insisting that
{\em all} pure states be uniquely identified by the expectation values
of the POVM elements.  I will say that such a POVM is PSI{ \em really}
complete:

{\bf Definition (PSI {\em really}-completeness).}  A pure-state 
informationally {\em really} complete (PSI{\em R}-complete) POVM on a 
finite-dimensional quantum system is
a POVM whose outcome probabilities are sufficient to determine any
pure state (up to a global phase).

\vspace{0.3cm}
In the third section of this note I will prove two theorems about the
number of elements necessary for a POVM to be PSI{\em R}-complete:

\vspace{0.3cm}\noindent
{\bf Theorem I} Let \{$E_i$\} be the elements of a PSI{\em R}-complete
POVM for a system of dimension $d$.  Then for any non-zero vector 
$|\phi \rangle$ in the
Hilbert space, there are at least $2d-1$ elements which do not
annihilate $|\phi \rangle$ (that is, with $E_{i}|\phi \rangle \neq 0 $).

\vspace{0.4cm}\noindent
Since a POVM which is I-complete for all states, 
whether pure or mixed, is {\it a fortiori} PSI{\em R}-complete, the 
conclusion of Theorem I holds for those POVMs also.

\vspace{0.4cm}\noindent
{\bf Theorem II} A PSI{\em R}-complete  rank-one POVM for a system of
dimension $d$ must have at least $3d-2$ elements.

\vspace{0.4cm}\noindent
Together with the rank-one PSI-complete POVM with $2d$ elements
presented in the second section, Theorem II shows that allowing failure
of state identification on a set of measure zero does decrease the 
minimum number of elements in a rank-one POVM, for all $d>2$.

There has been some interest in discussing I-complete
measurements utilizing ``mutually-unbiased bases'' (MUBs)\cite{I,WF,WW}.
Two orthonormal bases $\{ |e_{i}\rangle \}$ and $\{ |f_{j}\rangle \}$
for a space of dimension $d$ are mutually unbiased if, for all $i$ and
$j$, $| \langle e_{i}| f_{j} \rangle | ^{2} = 1/d $.  It is known
that, for some values of $d$, there exist $d+1$ MUBs, and
that in those cases the set of projectors on all of those basis elements 
is I-complete \cite {I}.  That would be a total of $(d+1)d = d^{2}+d$
projectors,
but because not all of the expectation values of these projectors are
independent, this set can be related to a rank-one POVM with $d^2$
elements, which is the minimum number of elements of a I-complete POVM
\cite {CFS}.  It is also known \cite {KR} that for any value of 
$d$ there does exist at least 3 MUBs (and it is conjectured
\cite {Z, G} that for some values of $d$ no more than that), which can
be related to a rank-one POVM with $3d-2$ elements.  This is the smallest
number not ruled out by Theorem II, and so one might hope that this POVM
could be PSI{\em R}-complete.  However, this would certainly not be true
if there were a fourth basis mutually-unbiased with respect to the other 
three, since the expectation values for any two elements of this fourth 
basis would coincide, and hence those two elements could not be uniquely
identified. 
Furthermore, while Theorem II establishes that any rank-one
PSI{\em R}-complete POVM has at least $3d-2$ elements, it does not
assert that a greater number might not in fact be required.  To my knowledge,
the minimum number of elements of a rank-one PSI{\em R}-complete POVM
is at present unknown.
   
\section{A rank-one PSI-complete POVM with \mbox{$2d$ elements}}
   
In this section I will show that, for any dimension $d$, there exists
a rank-one PSI-complete POVM with $2d$ elements, thus confirming the
conjecture made by FSC.

Let $\{ |e_{i}\rangle |i=0,\ldots ,d-1\} $ be an orthonormal basis for
a Hilbert space of dimension $d$.  Write a vector in this space as
\begin{equation} 
                |\psi \rangle = \sum_{i=0}^{d-1} c_{i}|e_{i}\rangle .
\end{equation}
I will use the global phase freedom, and the indifference to sets
of measure zero in the definition of PSI-completeness,
to assert that $c_0$ is real and strictly positive.
Now consider the following set of operators:\\
\indent The $d$ operators $|e_{i}\rangle \langle e_{i}| $, for
$i=0,\ldots ,d-1$, and \\
\indent The $(d-1)$ operators $(|e_{0}\rangle + i|e_{i}\rangle)
(\langle e_{0}| -i\langle e_{i}|)$ for $i=1,\ldots ,d-1$, and \\
\indent The single operator $ (\sum_{i=0}^{d-1} |e_{i}\rangle)
(\sum_{j=0}^{d-1} \langle e_{j}|)$.\\
This is a set of $2d$ operators, which I will show is PSI-complete; that 
is, for any vector $|\psi \rangle$ outside of a set of measure zero,
knowledge of the expectation values of these operators would enable one
to calculate the values of the coefficients $c_i$.   

We have the expectation values
\begin{equation} \langle \psi | (|e_{i}\rangle \langle e_{i}|)|\psi \rangle
               = |c_{i}|^2 ;
\end{equation}
this gives us the value of $|c_{i}|$ for $i=1,\ldots ,d-1$, and (since
$c_{0} > 0$) the value of $c_0$.  For $i=1,\ldots ,d-1$, we also have the 
expectation values
\begin{equation} \langle \psi |  
         (|e_{0}\rangle + i|e_{i}\rangle)
         (\langle e_{0}| -i\langle e_{i}|)|\psi \rangle  
          = c_{0}^{2} + |c_{i}|^{2} + 2c_{0}\Im c_{i};
\end{equation}
together with the values of $c_0$ and of $|c_{i}|$, this gives us the 
value of $\Im c_{i}$. And since
$(\Re c_{i})^{2} = |c_{i}|^{2} - (\Im c_{i})^{2}$,
at this point we know everything except for the signs of $(\Re c_{i})$ 
for $i=1,\ldots ,d-1$.

We still have one more expectation value, namely
\begin{eqnarray} \langle \psi | (\sum_{i=0}^{d-1} |e_{i}\rangle)
    (\sum_{j=0}^{d-1} \langle e_{j}|)|\psi \rangle & = & \nonumber \\
    |\sum_{i=0}^{d-1} c_{i}|^{2} & = & (\sum_{i=0}^{d-1} \Re c_{i})^{2} +
    (\sum_{i=0}^{d-1} \Im c_{i})^{2},
\end{eqnarray}
and so we know the value of $|(\sum_{i=0}^{d-1} \Re c_{i})|$. I will show
that, in the generic case, this is enough to tell us the sign of each
$\Re c_i$, and hence to uniquely identify $|\psi \rangle $.
Suppose for example that we knew that $c_{0} = +5$, $|\Re c_{1}| = 8$,
$|\Re c_{2}| = 4$, and that $|c_{0}+\Re c_{1}+\Re c_{2}|=7$; 
this would tell us
that $\Re c_{1} = -8$ and that $\Re c_{2} = -4$.  To see in general what 
ambiguities are allowed by all of the expectation values, suppose that
a given set of expectation values were compatible with
two distinct vectors $|\psi \rangle $ and $|\psi ^ \prime \rangle$,
with coordinates $c_i$ and $c_{i}^\prime $ respectively. It would then
be true that $c_{0} = c_{0}^{\prime } $, that $|\Re c_{i}| =
|\Re c_{i}^{\prime }|$ for $i=1,\ldots ,d-1$, 
and that $|(\sum_{i=0}^{d-1} \Re c_{i})| =
|(\sum_{i=0}^{d-1} \Re c_{i}^{\prime}) |$.
I will divide the coordinates into two sets, according to whether
$\Re c_{i} = + \Re c_{i}^\prime $ or $\Re c_{i} = - \Re c_{i}^\prime $.
Define $E := \{ i|\Re c_{i} = \Re c_{i}^{\prime} \} $; since 
$\Re c_{0} = \Re c_{0}^\prime $, $E$ is not empty.  Also define
$U := \{ i|\Re c_{i} \neq \Re c_{i}^{\prime} \} $; note that 
$\Re c_{i} = -\Re c_{i}^\prime $ for $i \in U$, and that, since
$|\psi \rangle $ and $|\psi ^{\prime} \rangle $ were assumed to be distinct,
$U$ is not empty. We know that either $(\sum_{i=0}^{d-1} \Re c_{i}) = +
(\sum_{i=0}^{d-1} \Re c_{i}^\prime) $ or $(\sum_{i=0}^{d-1} \Re c_{i}) = 
 -(\sum_{i=0}^{d-1} \Re c_{i}^\prime) $; if $(\sum_{i=0}^{d-1} \Re c_{i}) 
= +(\sum_{i=0}^{d-1} \Re c_{i}^\prime) $, then 
\begin{equation} \sum_{i\in E} \Re c_{i} +\sum_{i\in U} \Re c_{i} =
    \sum_{i\in E} \Re c_{i}^{\prime} +\sum_{i\in U} \Re c_{i}^{\prime} .
\end{equation}
which gives
\begin{equation}\sum_{i\in E} \Re c_{i} +\sum_{i\in U} \Re c_{i} =
    \sum_{i\in E} \Re c_{i} -\sum_{i\in U} \Re c_{i}
\end{equation}
and so
\begin{equation}\sum_{i\in U} \Re c_{i} = 0.
\end{equation}
On the other hand, if $(\sum_{i=0}^{d-1} \Re c_{i}) = 
-(\sum_{i=0}^{d-1} \Re c_{i}^\prime) $, then
\begin{equation} \sum_{i\in E} \Re c_{i} +\sum_{i\in U} \Re c_{i} =
    -(\sum_{i\in E} \Re c_{i}^\prime +\sum_{i\in U} \Re c_{i}^{\prime}) .
\end{equation}
which gives
\begin{equation}\sum_{i\in E} \Re c_{i} +\sum_{i\in U} \Re c_{i} =
    -(\sum_{i\in E} \Re c_{i} -\sum_{i\in U} \Re c_{i}),
\end{equation}
and so
\begin{equation}\sum_{i\in E} \Re c_{i} = 0.
\end{equation}
So either eq.\ 7 or eq.\ 10 must be correct.  For fixed sets $E$ and $U$, 
states satisfying either of these equations are confined to  closed sets
of measure zero.  And since there are only a finite number of 
possibilities for $E$ and $U$, all states outside of a closed set of 
measure zero can be unambiguously identified, and so our operators are
PSI-complete.
 
Given these rank-one PSI-complete operators, we can form a rank-one
PSI-complete POVM with $2d$ elements with the same procedure used by
FSC: denoting our operators by $P_i$, with $i=1,\ldots ,2d$, define
$G=\sum_{i=1}^{2d}P_i$; this is non-singular since the $P_i$ are 
PSI-complete, so we can define 
$E_{i} := G^{-\frac{1}{2}}P_{i}G^{-\frac{1}{2}}$.  These operators are
then the $2d$ elements of a rank-one PSI-complete POVM.

\section{Proof of the theorems}
In this section I will prove the two theorems stated in the Introduction.

\vspace{0.5cm}\noindent
{\bf Theorem I} Let \{$E_i$\} be the elements of a PSI{\em R}-complete
POVM for a system of dimension $d$.  Then for any non-zero vector 
$|\phi \rangle$ in the
Hilbert space, there are at least $2d-1$ elements which do not
annihilate $|\phi \rangle$ (that is, with $E_{i}|\phi \rangle \neq 0 $).

\vspace{0.5cm}\noindent
Proof: Let \{$E_i$\} be the elements of a POVM for a system of dimension 
$d$. For any vector $|\phi \rangle $, let $K_\phi $ be the
number of POVM elements with $E_{i}|\phi \rangle \neq 0 $.
I will show that if there is a non-zero vector with $K_\phi < 2d-1$ then 
this POVM is not PSI{\em R}-complete.

Let $|\phi \rangle $ be a non-zero vector with $K_\phi < 2d-1$,
and let $|\chi \rangle $ denote a non-zero vector orthogonal to
$|\phi \rangle $.  I will show that $|\chi \rangle $ can be chosen
so that
\begin{equation} \Re \langle \phi |E_{i} |\chi \rangle = 0
\end{equation}
for all elements $E_i$.  For those $E_i$ which annihilate 
$|\phi \rangle$, eq.\ 11 is valid for any $|\chi \rangle $, so eq.\ 11
represents $K_\phi $ conditions which $|\chi \rangle $
must satisfy. These conditions are not all independent; since 
$\sum_{i}E_{i}= I$ implies that
\begin{equation}\sum_{i}\Re \langle \phi |E_{i} |\chi \rangle
                =\Re \langle \phi |\chi \rangle = 0,
\end{equation}
there are not more than $K_{\phi}-1$ independent conditions.
Now let $\{ |e_{j}\rangle |j=1,\ldots ,d-1 \} $ be an orthonormal basis
for the subspace orthogonal to $|\phi\rangle $; then for each value
of $i$, eq.\ 11 can be written
\begin{equation}  \sum_{j=1}^{d-1}[\Re \langle \phi |E_{i}|e_{j}\rangle
   \Re \langle e_{j}|\chi \rangle -\Im \langle \phi |E_{i}|e_{j}\rangle
   \Im \langle e_{j}|\chi \rangle] = 0.
\end{equation}
For each value of $i$, this is a real, linear homogeneous equation
for the $2d-2$ real parameters $\Re \langle e_{j}|\chi \rangle$
and $\Im \langle e_{j}|\chi \rangle$.  Since no more than $K_{\phi}-1$
of these equations are independent, and since  $K_{\phi}-1<2d-2$,
there must be a non-trivial solution.

So we can choose $|\chi \rangle $ to satisfy eq.\ 11, and then define
\begin{equation} |\psi_{\pm} \rangle = |\phi \rangle \pm |\chi \rangle .
\end{equation}
The expectation values are
\begin{equation} \langle \psi_{\pm}|E_{i}|\psi_{\pm}\rangle  = 
                 \langle \phi |E_{i}|\phi \rangle +
                  \langle \chi |E_{i}|\chi \rangle \pm
                  2\Re \langle \phi |E_{i}|\chi \rangle 
\end{equation}
Eqs.\ 11 and 15   together imply that
\begin{equation}\langle \psi_{+}|E_{i}|\psi_{+}\rangle  =
                 \langle \psi_{-}|E_{i}|\psi_{-}\rangle  
\end{equation}
for all $i$.  Finally, let $N$ be the (common) norm of 
$|\psi_{+}\rangle $ 
and of $|\psi_{-}\rangle $; then the normalized vectors 
$(N^{-1}) |\psi_{\pm}\rangle $ 
represent two distinct states whose
expectation values for all of the POVM elements agree, and so the POVM 
is not PSI{\em R}-complete.\footnote[2]{I could of course satisfy
eq.\ 11 with much less work by simply taking either $|\chi \rangle $
or $|\phi \rangle $ to be $0$.  However, if $|\chi \rangle $ were $0$
the states $|\psi_{+}\rangle $ and $|\psi_{-}\rangle $ would be equal; if 
$|\phi \rangle $ were $0$, then $|\psi_{+}\rangle =
- |\psi_{+}\rangle $, and the fact
that their expectation values agreed would follow from the global 
phase ambiguity, and so would not imply that the POVM failed to be
PSI{\em R}-complete.} 

\vspace{0.5cm}\noindent
{\bf Theorem II} A rank-one PSI{\em R}-complete POVM for a system of
dimension $d$ must have at least $3d-2$ elements.

\vspace{0.5cm}\noindent
Proof: Let $\{ E_{i} \} $ be the elements of a rank-one 
PSI{\em R}-complete POVM for a system of dimension $d$.  Define 
$F := \sum_{i=1}^{d-1}E_i$; since each $E_i$ has rank one, the rank of
$F$ is no greater than $d-1$.  This means that there must be (at least)
a one-dimensional subspace annihilated by $F$, and hence, since the
POVM elements are positive, by each $E_i$ for $i=1,\ldots ,d-1$.
Let $|\phi \rangle $ be any non-zero vector in that subspace.  
$|\phi \rangle $
is annihilated by each $E_i$ for $i=1,\ldots ,d-1$; in addition, according
to theorem I there must be at least $2d-1$ elements which do {\em not}
annihilate $|\phi \rangle $.  Therefore the POVM must contain at least
$(d-1)+(2d-1)=3d-2$ elements. 

\vspace{1cm}
{\bf Acknowledgments:} I would  like to acknowledge
correspondence with Carl Caves and with Steve Flammia, and also the 
hospitality of the Lawrence Berkeley National Laboratory, where this 
work was done.
\end{document}